\documentstyle[twocolumn,epsfig,prb,aps,amsmath,amssymb]{revtex}
\begin{document}

\twocolumn[\hsize\textwidth\columnwidth\hsize\csname
@twocolumnfalse\endcsname

\title{Effects of interladder couplings in the trellis lattice}

\author{Jos\'e A. Riera and Sergio D. Dalosto}
\address{
Instituto de F\'{\i}sica Rosario, Consejo Nacional de 
Investigaciones 
Cient\'{\i}ficas y T\'ecnicas, y Departamento de F\'{\i}sica,\\
Universidad Nacional de Rosario, Avenida Pellegrini 250, 
2000-Rosario, Argentina}
\date{\today}
\maketitle
\begin{abstract}
Strongly correlated models on coupled ladders in the presence of
frustration, in particular the trellis lattice, are studied by
numerical techniques. For the undoped case,
the possibility of incommensurate peaks in the magnetic
structure factor at low temperatures is suggested. In the
doped case, our main conclusion for the trellis lattice is that by
increasing the interladder coupling, the balance between the magnetic
energy in the ladders and the kinetic energy in the zig-zag chains
is altered leading eventually to the destruction of the hole pairs
initially formed and localized in the ladders.
\end{abstract}


\vskip2pc]
\section{Introduction}
                 
The study of strongly correlated systems in low spatial dimensions
is nowadays the center of an intense effort both theoretically and
experimentally. Among these
low-dimensional systems the ones containing two-leg ladders have
received considerable attention. One of the original motivation for
the study of ladders was the search for a simple mechanism for pairing
in a strongly correlated model.\cite{drs} This possibility was
later confirmed by a number of studies and the symmetry of the
superconducting order parameter was found to be
d$_{x^2-y^2}$ (Refs.~\onlinecite{rice93,noack}).
A considerable interest in these theoretical predictions was renewed
by the discovery of superconductivity in the ladder compound
Sr$_{14}$Cu$_{24}$O$_{41}$ (14-24-41) after being doped with Ca and
under a pressure of $\approx 3 \rm GPa$ (Ref.~\onlinecite{uehara}).

However, the 14-24-41 compound, as well as many other compounds,
like another cuprate SrCu$_2$O$_3$ and the vanadates CaV$_2$O$_5$
and NaV$_2$O$_5$, actually contain layers of  two-leg ladders which
are {\em coupled} by frustrated effective interactions in the
so-called trellis lattice.
The strength of these frustrating interladder couplings may
be weak enough to consider the ladders as essentially isolated or
strong enough to change radically the physical behavior of a single
ladder.

The original experiments on Sr$_{0.4}$Ca$_{13.6}$Cu$_{24}$O$_{41}$
(Ref.~\onlinecite{uehara}) reveal that the superconducting
critical temperature reaches its maximum of $\rm 12 K$ at a
pressure of $\approx 3 \rm GPa$ and then decreases as the pressure
is further increased. Similar results were obtained for
Sr$_{2.5}$Ca$_{11.5}$Cu$_{24}$O$_{41}$ (Ref.~\onlinecite{nagata}).
The application of pressure to this compound may change the strength
of some couplings, or to additionally increase the doping of the
ladder layers as holes are transfered from the chain layers also
present in this compound. The main purpose of this paper is to
analyze the first possibility, neglecting more radical changes
in the crystallographic structure.\cite{pachot}
In a highly simplified model for this compound, we consider three
sets of couplings: along the ladder legs, along the rungs, and on
the ``zig-zag" interactions between the ladders.\cite{Note1} The
effect of varying the rung interactions, keeping fixed the leg ones,
on a single ladder, has been analyzed extensively.\cite{rpd}
In the present study we will concentrate on the effects of varying
the interladder couplings specially on the magnetic and
pairing properties.

To this purpose we study the $t-J$ model which is appropriate to
describe these cuprates and vanadates characterized by large on-site
Coulomb repulsion and close to half-filling. We start our studies
with undoped furstrated coupled ladders. For the trellis latttice,
it was suggested that it is possible a transition
from a spin liquid to a possible spiral order with incommensurate
magnetic correlations as the interladder coupling (ILC)
increases.\cite{normandmila} Then, we will consider specially the
case of two-hole doping, i.e. the evolution of pairing as the
interladder coupling is varied. Preformed hole pairs are already
present in the uncoupled ladders\cite{drs,noack,magishi} and a small
ILC could lead to superconductivity (SC) as a proximity effect
between the ladders. However, our main concern in this work is not
the onset of SC but rather the effect on pairing due to somewhat
large ILC.

There are further motivations to study both experimentally the
14-24-41 compound and theoretically the $t-J$ model on the trellis
lattice. It is in effect remarkable how a relatively small difference
like the one between the square lattice of the Cu-O planes in
high-T$_c$ cuprates and the trellis lattice in 14-24-41 leads to
such a considerable difference in the superconducting properties of
those materials. This difference is even more remarkable if we take
into account the presence of stripes~\cite{tranquada} in the
underdoped regime and (at least) above the superconducting region of
the high-T$_c$ cuprates. These stripes can be thought as metallic
ladders separated by insulating antiferromagnetic ones, specially in
the ``bond-centered" stripes obtained from a numerical study of the 2D
t-J model\cite{white}.

In this sense, the $t-J$ model on the trellis lattice is a testing
ground for the study of the competition between magnetic and kinetic
energies which is at the core of the mechanism of micro-phase
separation leading to the formation of stripes in the high-T$_c$
cuprates.\cite{white} Our main conclusion for the trellis lattice is
that by increasing the interladder coupling, the balance between the
magnetic energy in the ladders and the kinetic energy in the zig-zag
chains in between the ladders
is altered leading eventually to the destruction of the hole pairs
initially formed and localized in the ladders. We also suggest the
possibility that the hole pairs may go to the zig-zag chains
in a process which represents a transition from a magnetic to a
kinetic mechanism of pair binding.

\begin{figure} 
\begin{center}
\epsfig{file=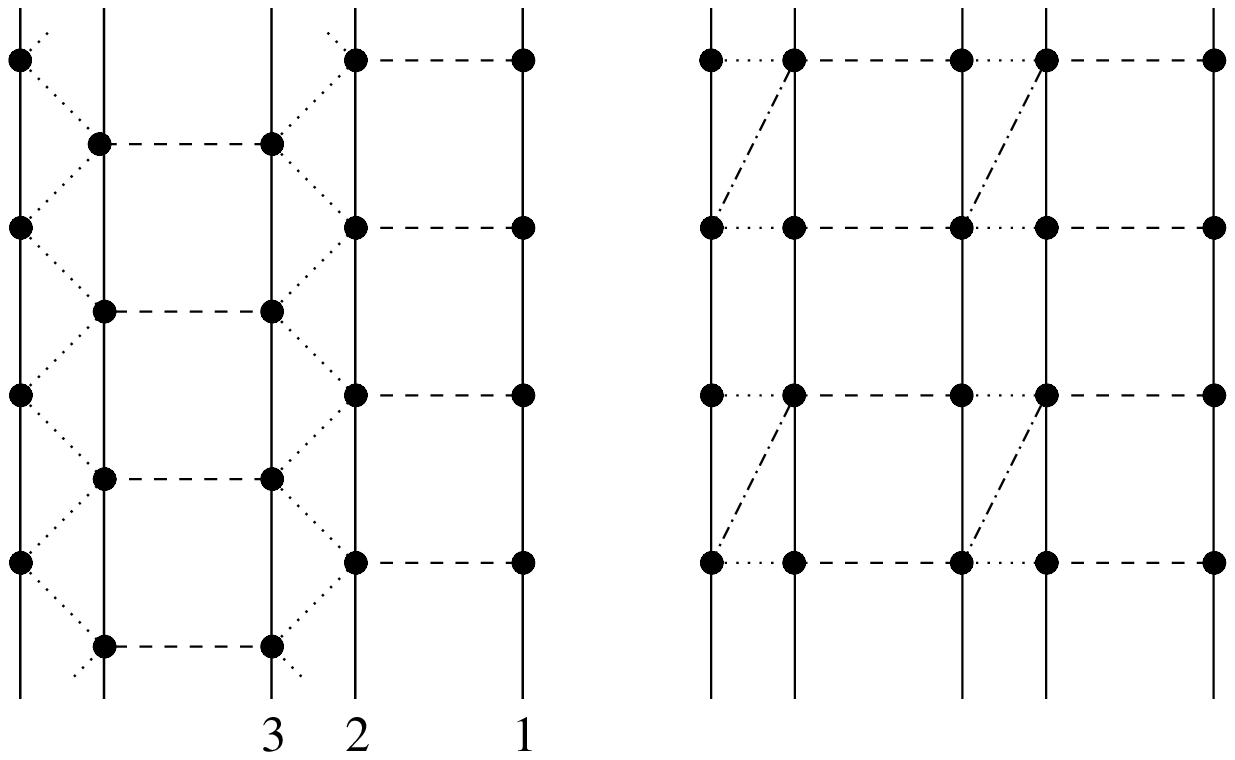,width=8cm,angle=0}
\end{center}
\caption{Left: trellis lattice. Right:
modified trellis lattice. Ladder legs and rungs are indicated 
with full and dashed lines respectively. In the modified lattice,
``perpendicular" and ``diagonal" interladder couplings
are indicated with doted and dash-dotted lines respectively.
}
\label{trelfig}
\end{figure}

The t-J model on the trellis lattice is given by the Hamiltonian:
\begin{eqnarray}
{\cal H}={\cal H}_{leg} + {\cal H}_{rung} + {\cal H}_{inter}
\label{hamtrel}
\end{eqnarray}
\noindent
where
\begin{eqnarray}
{\cal H}_\alpha =&-&t_\alpha \sum_{ \langle i j
\rangle,\sigma }({\tilde c}^{\dagger}_{i \sigma}
{\tilde c}_{ j \sigma} + h.c. ) \nonumber \\
&+&~J_\alpha \sum_{
\langle i j \rangle }( {\bf S}_{ i} \cdot
{\bf S}_{ j} -{\frac{1}{4}} n_{ i} n_{ j} )
\nonumber
\end{eqnarray}
The couplings are $(t,J)$, $(t^\prime,J^\prime)$ and
$(t_{inter},J_{inter})$ along the legs, along the rungs and between
the ladders respectively (Fig.~\ref{trelfig}). The rest of the
notation is standard.
Periodic boundary conditions in both directions are considered
except otherwise stated. We adopt $J=0.4 t$, a value usually taken to
model high-T$_c$ cuprates, and in order to reduce the number of
independent variables $J_\alpha=J(t_\alpha/t)^2$. Moreover, we
take $t=1$.
For the undoped compound, neutron scattering experiments for the
14-24-41 compound\cite{eccleston} suggest $J=2 J^\prime$. However,
taking into account the analysis mentioned in Ref~.\onlinecite{Note1}
most of our calculations have been done for the isotropic case.

We use various numerical techniques like quantum Monte Carlo (QMC),
with a conventional world-line algorithm, and exact diagonalization
with the Lanczos algorithm (LD), complemented by the continued
fraction formalism to compute dynamical properties.

\section{Frustrated coupled spin ladders}
\label{undoped}

\begin{figure} 
\begin{center}
\epsfig{file=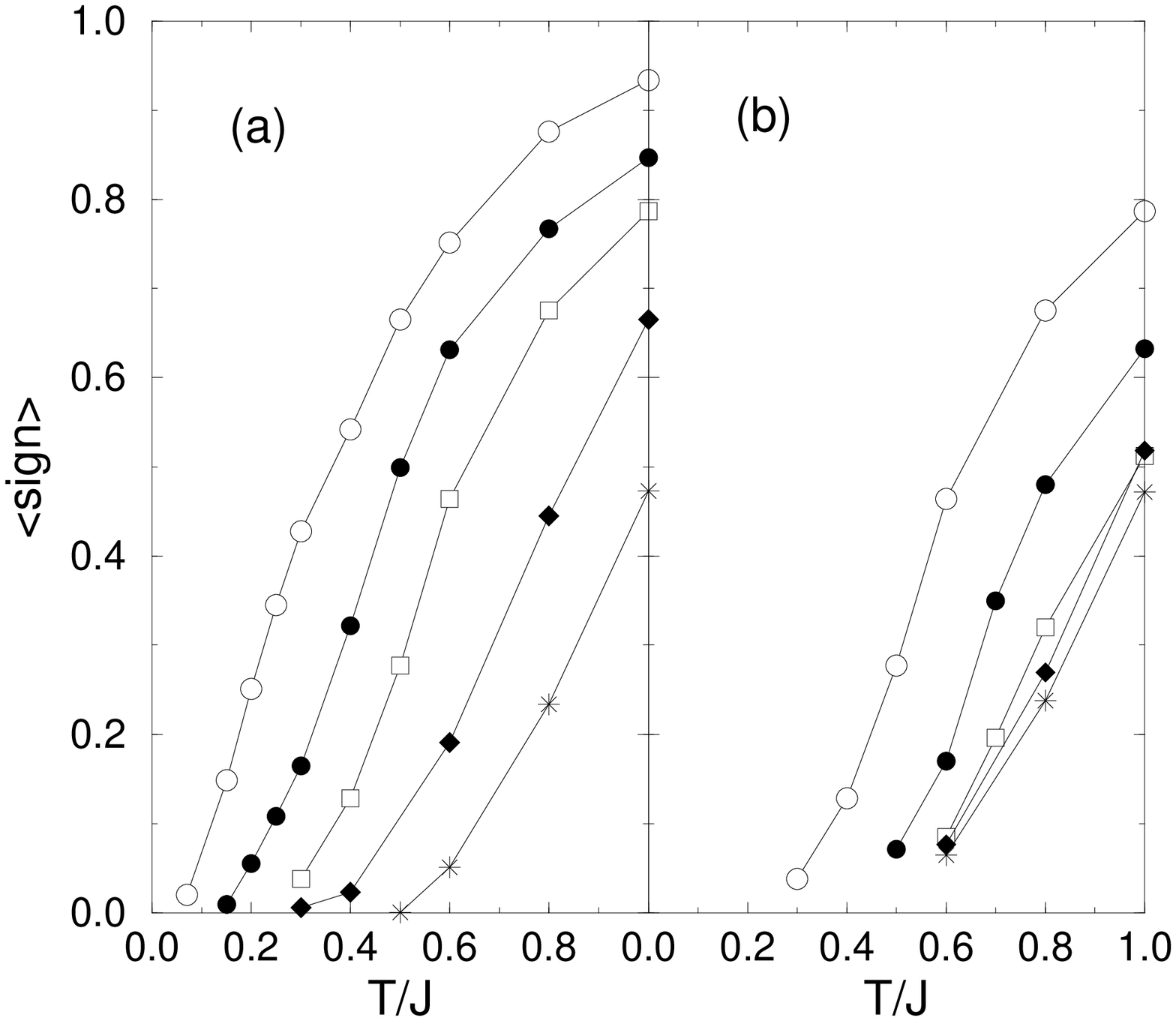,width=7cm,angle=-90}
\end{center}
\caption{
Average sign of the QMC on the modified trellis lattice,
(a) at $J=J^{\prime}=1$, $J_{perp}=J_{diag}=0.2$ for
$4\times4$ (open circles), $6\times6$ (full circles), $8\times8$
(squares), $12\times12$  (diamonds) and $16\times16$ (stars)
and (b) for the $8\times8$ lattice, $\rm J=J^\prime=1$, 
$J_{perp}=0.4$ and $J_{diag}=0.4$ (open circles), 0.8 (full
circles), 1.2 (squares), 1.6 (diamonds), 2.0 (stars).
}
\label{signo}
\end{figure}

The first issue we want to address is the evolution of the magnetic
order in the absence of doping as the interladder coupling is
increased. To this end we have computed the static magnetic structure
factor $\rm S(\bf q)$ on $\rm L \times L$ clusters using QMC. Due to
the presence of frustration, the minus sign problem prevents us to
reach low temperatures and to study larger clusters which would be
necessary to perform a finite size scaling. The same problems were
already faced in a previous QMC study of the susceptibility of this
system.\cite{miyahara} In order to reduce the minus sign problem
and to take advantage of the simple checkerboard
decomposition,\cite{reger}
we take a slightly modified lattice in which the interladder
couplings are ``perpendicular" ($J_{perp}$) and ``diagonal" 
($J_{diag}$) as shown in Fig.~\ref{trelfig}.
This lattice contains just half of the diagonal interladder couplings
than in the trellis lattice, and we have checked by exact
diagonalization on small clusters that this difference does not
change qualitatively the results.
Even for this modified lattice the minus sign problem 
is severe as shown in Fig.~\ref{signo}. We recall that the average
of any observable ${\cal O}$, in a system which presents this problem
is computed as:\cite{wiese}
\begin{eqnarray}
\langle {\cal O}\rangle= \frac{\langle {\cal O} Sign\rangle}
{\langle Sign \rangle}
\nonumber
\end{eqnarray}
\noindent
with respect to a modified partition function of the $2+1$-dimensional
problem, ${\cal Z'}= \sum_s |\exp{S(s)}|$
where $S(s)$ is an effective action. In particular, the average
sign is:
\begin{eqnarray}
{\langle Sign \rangle } = \frac{1}{\cal Z'} \sum_s Sign(s) |\exp{S(s)}|,
\nonumber
\end{eqnarray}
\noindent
where $Sign(s)=sign(\exp{S(s)})$. Then ${\langle Sign \rangle }$ 
is the ratio of the original partition function
${\cal Z}= \sum_s \exp{S(s)}$ to ${\cal Z'}$.
In practice, at each measurement step, $\exp{S(s)}$ is computed as
the product of the transition elements of

\noindent
all the cubes that
makes up the $2+1$-dimensional lattice.\cite{reger}
Nonetheless, although we cannot compute some quantities in
the bulk limit, we can indicate qualitatively the behavior of the
magnetic order as a function of $J_{inter}$ (we take $J$ as the
unit of energies in this section). 

\begin{figure} 
\begin{center}
\epsfig{file=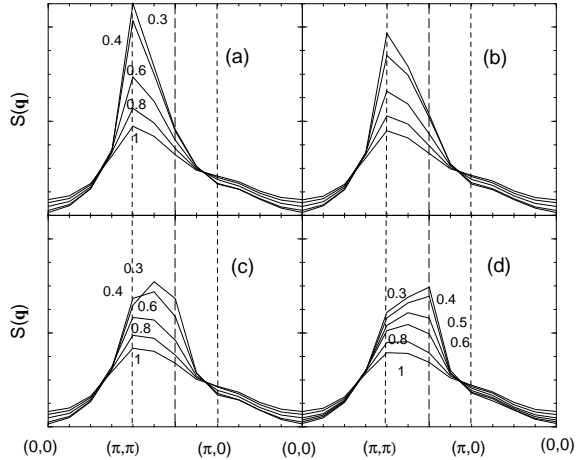,width=6cm,angle=-90}
\end{center}
\caption{
Magnetic static structure factor $\rm S(\bf q)$ obtained by QMC on
a $8\times8$ cluster of the modified trellis lattice, at several
temperatures (in units of $\rm J$), $\rm J=J^\prime=1$, 
$J_{perp}=0.2$ and (a)
$J_{diag}=0.2$, (b) 0.4, (c) 0.6 and (d) 0.8.
}
\label{stfac}
\end{figure}

Typical results are shown in Fig.~\ref{stfac} for coupled isotropic
($J^\prime=J$) ladders on the $8\times 8$ cluster.
In  Fig.~\ref{stfac}(a) we show for $J_{perp}=J_{diag}=0.2$ the
characteristic structure factor of isolated ladders with a peak at
$(q_x,q_y)=(\pi,\pi)$ ($x$ ($y$) is the direction along (transversal)
to the ladders). This peak becomes more pronounced as the temperature
is lowered. On the other hand, keeping $J_{perp}=0.2$ and as 
$J_{diag}$ is increased, the peak
starts to shift from $(\pi,\pi)$ to $(\pi,\pi/2)$
(Fig.~\ref{stfac}(b,c,d)). A second interesting feature should be
noticed in Fig.~\ref{stfac}(d): the peak of $\rm S(\bf q)$ is located
at $(\pi,\pi)$ at high temperature (in this case down to $\rm T \approx
0.8$, in units of $J$) and as the temperature is lowered
it starts to shift away from $(\pi,\pi)$. At low temperatures (in this
case below $\rm T \approx 0.4$ the peak is located at $(\pi,\pi/2)$.
Since it is clear that this behavior is caused by the frustration of
the interladder couplings, it will certainly be present in the
original trellis lattice. As indicated in Fig.~\ref{stfac}c, it is 
possible that an incommensurate peak across the ladder direction
could be present at intermediate values of $J_{diag}$ and 
intermediate temperatures.

The second point we want to examine is the behavior of the excitations
of these systems, in particular the $S=1$ excitations as can be 
measured by neutron scattering experiments. For this purpose, using
conventional LD with the standard continued fractions
formalism,\cite{haas} we have computed the zero temperature dynamical
structure function ($zz$ component) $S({\bf q},\omega)$.
In this case, we have to
limit ourselves to somewhat smaller clusters but we are confident that
the qualitative features we found will survive in the bulk limit.

\begin{figure} 
\begin{center}
\epsfig{file=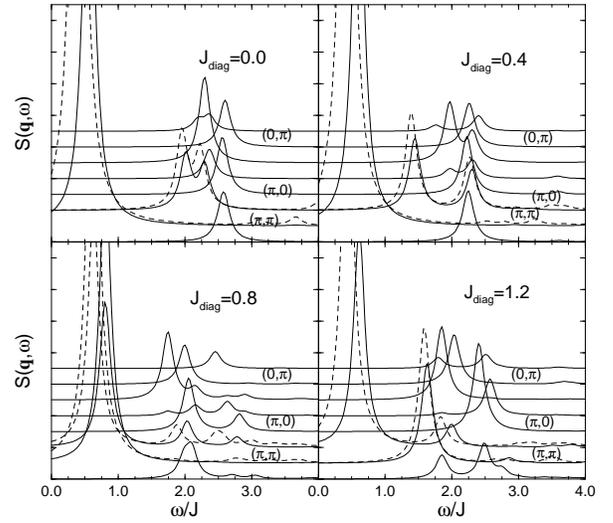,width=7cm,angle=-90}
\end{center}
\caption{
Dynamical structure factor $S({\bf q},\omega)$ for several momentum
for $\rm J$), $\rm J=J^\prime=1$, $J_{inter}=0.8$ and various
values of $J_{diag}$ as indicated in the figures.
These results were obtained on a system of two coupled 
$2\times4$ (full lines) and $2\times6$ (dashed lines) ladders.
}
\label{dynamic}
\end{figure}

Results obtained for the $4 \times 4$ cluster are shown in
Fig.~\ref{dynamic}.
In the absence of frustration (Fig.~\ref{dynamic}(a)) the peak in
$S({\bf q},\omega)$ which corresponds besides to the lowest
excitation, is located at $(\pi,\pi)$, as expected in the bulk
limit for an AF order. As a frustrating ILC is increased
(Fig.~\ref{dynamic}(b,c,d)) it can be seen that considerable 
spectral weight is transferred to the peak at $(\pi,\pi/2)$,
which becomes finally the lowest energy excitation. Similar results
are also shown for the $4 \times 6$ cluster for ${\bf q}=(\pi,\pi)$
and $(\pi,\pi/2)$.

The results shown in Figs.~\ref{stfac} and ~\ref{dynamic} are
unequivocally due to frustration and are qualitatively similar to the
ones previously obtained in a system of ferromagnetically (FM)
coupled ladders.\cite{dalosto} Similar results have been obtained by
LD on the $4\times4$ and $4\times6$ clusters of the real trellis
lattice. In the case of the trellis lattice, as in the FM ILC case,
we expect that the behavior above discussed will be present in the
bulk limit for strong enough interladder couplings and low enough
temperatures. The impossibility of assessing finite size effects
prevents us to determine if this behavior is present for arbitrarily
small values of $\rm J_{inter}$ or, on the contrary, only for values
larger than a ``critical" one.

We have not detected any sign of
incommensurability along the ladder direction.\cite{normandmila}
Such an incommensurability is expected in principle since a trellis
lattice can be also considered as coupled $J_1-J_2$ chains
($J_1=J_{inter}$, $J_2=J$, in our notation) which are known to
present a peak in $S(q)$ at a momentum which continuously
varies from $\pi$ to $\pi/2$ (as defined on our modified trellis
lattice) as $J_2 / J_1$ goes from $\infty$ to
zero.\cite{whiteaffleck} However, notice that, as can be seen in
Fig.~\ref{signo}, values of $J_{inter} > 0.6$ at low enough
temperatures cannot be reached in our simulations.\cite{Note3}

\section{Doped trellis lattice.}
\label{doped}

We now analyze the hole pairing in the doped trellis lattice as the
interladder couplings are increased. To gain some insight in this
problem we start by considering an isolated building block of
the trellis lattice. This minimal system is a three chain cluster
consisting of a ladder and a zig-zag chain, (lines ``1", ``2" and
`3' in Fig.~\ref{trelfig}). The Hamiltonian is the one defined in
Eq.~(\ref{hamtrel}). All the results in this section have been
obtained by exact diagonalization. The justification of this study
involving somewhat small clusters is based on an extensive body of
similar studies of strongly correlated models which shows that
an important part of the physics of these models is dominated by
short range effects, which are appropriately captured in these
small cluster calculations.

\begin{figure}
\begin{center}
\epsfig{file=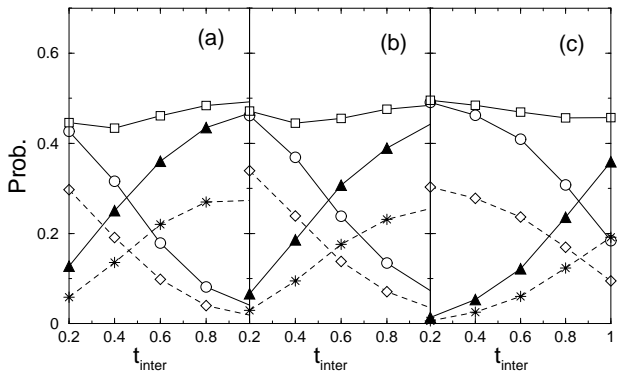,width=4.2cm,angle=-90}
\end{center}
\vspace{0.5cm}
\caption{
Probability of finding a hole in the outer (circles) and inner
(squares) ladder legs, and in the third chain (triangles) of the
$3\times 6$ cluster with two holes for (a) $t^\prime=0.75$,  (b)
$t^\prime=1.0$ and (c) $t^\prime=1.5$. The probability of
finding a hole in chain ``1" and the other hole in chain ``2",
and the probability of finding a hole in chain ``2"
and the other in chain ``3" are indicated with
diamonds and stars respectively.
}
\label{holeleg}
\end{figure}

In
this minimal trellis lattice already appears, upon doping with two
holes, the main feature we want to emphasize. In 
Fig.~\ref{holeleg}, the relative hole occupancy (or probability
of finding a hole) on each chain is shown in the $3\times 6$ cluster
with two holes
for $t^\prime=0.75$, 1.0 and 1.5 as a function of $t_{inter}$.
In the three cases, for small $t_{inter}$ the holes are almost
completely located in the ladder legs. As $t_{inter}$ increases
the probability of finding a hole in the outer ladder leg
decreases while increases the occupancy of chain ``3" which
is connected to the ladder by the zig-zag interaction. There
is a neat change of behavior, from a situation in which the
occupancy of chain ``3" is virtually unoccupied to a situation
in which the unoccupied chain is ``1". moreover, this crossover
is rather abrupt, specially for $t^\prime=0.75$.

For relatively small $t_{inter}$, the hole distribution is
typical of that of an isolated ladder\cite{rieradag}, i.e. they
form a bound pair with one hole on each leg.
On the other hand, for larger
values of $t_{inter}$ ($t_{inter} > 0.6$ for $t^\prime=1.0$
and $t_{inter} > 0.9$, for $t^\prime=1.5$, on the $3\times 8$
cluster) the holes have moved from the ladder to the two chains
containing the zig-zag interactions. As we indicated in the
previous section, these two chains with the
zig-zag interactions form a chain with first and second neighbor
interactions, in this case with $t-J$ couplings, i.e. a
frustrated $t_1-t_2-J_1-J_2$ chain.\cite{ogata}

\begin{figure}
\begin{center}
\epsfig{file=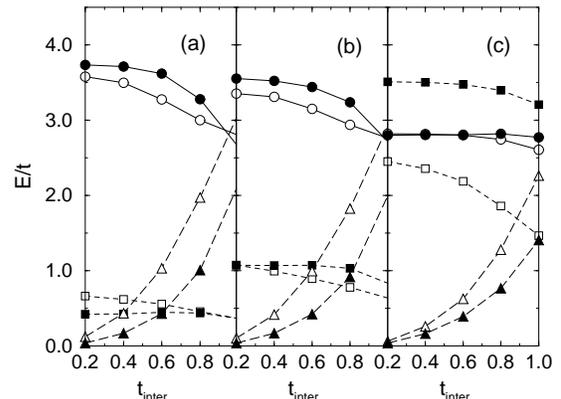,width=4.5cm,angle=-90}
\end{center}
\vspace{0.5cm}
\caption{
Energy contributions from different terms of the Hamiltonian
along the legs (circles), rungs (squares) and zig-zag (triangles).
Full (empty) symbols correspond to magnetic (kinetic) energies
in absolute value. Results obtained on a $3 \times 6$ cluster
for (a) $t^\prime=0.75$, (b) $t^\prime=1.0$ and (c) $t^\prime=1.5$.
}
\label{contr3x8h2}
\end{figure}

To understand the mechanism that produces this change in hole
pairing, let us
examine the contributions to the total energy from different
terms of the Hamiltonian (\ref{hamtrel}) as the interladder
hopping is increased. As it can be seen in Fig.~\ref{contr3x8h2},
the main differences between the contributions from the ladder
and those of the zig-zag chain are (i) in the ladder the
magnetic energy dominates (along the legs for $t^\prime=1.0$ or
along the rungs for $t^\prime=1.5$) while in the zig-zag
the kinetic energy is the most important, and (ii) the
main gain in energy as $t_{inter}$ is increased comes from
precisely the hopping term of the zig-zag ILC while the
magnetic energy on ladders is the most strongly decreased.

\begin{figure}
\begin{center}
\epsfig{file=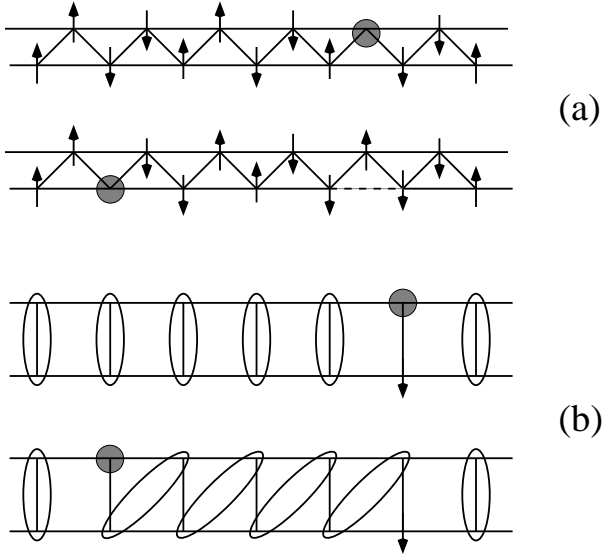,width=8cm,angle=0}
\end{center}
\caption{Schematic picture of the movement of a hole (a) in a
frustrated $t-J$ chain and (b) in a $t-J$ ladder. In each case,
the original (final) position of the hole is shown in the top
(bottom) panel.
}
\label{esquema}
\end{figure}

The gain in kinetic energy on the frustrated chain with respect to
the ladder can be explained by qualitative arguments as is
shown schematically in Fig~\ref{esquema}. In the frustrated
chain we have assumed an AF order of the spins along the chains
which is expected for $J_2 > J_1$ ($t_2 > t_1$). When the hole
moves, as in a simple $t-J$ chain, the hole leaves behind just
a single frustrated (ferromagnetic) bond. Something similar occurs
in the case of AF order along the zigzag chain ($J_1 \ge J_2$).
In the case of ladders, we have assumed a magnetic background
formed by spin singlets on the rungs. As a hole moves from its
initial position, it leaves behind a string of higher energy
singlets on the diagonals of the plaquettes. Hence there

\noindent
is a cost in energy which increases
roughly linearly with the distance traveled by the hole.

\begin{figure}
\begin{center}
\epsfig{file=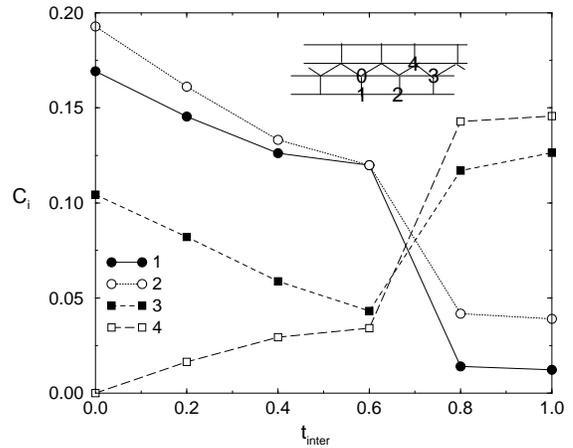,width=6cm,angle=-90}
\vspace{0.2cm}
\end{center}
\caption{Hole-hole correlation functions vs. $t_{inter}$ 
on the $4 \times 4$ cluster for $t^\prime=1.0$. The relative
distances from the reference site are indicated in the plot.
}
\label{hh4xn}
\end{figure}

What are the possible consequences of this behavior found in the
three-chain cluster for the trellis lattice? In this case, of
course, any chain along the ladder direction belongs at the same
time to a ladder and to a frustrated interladder chain. The
question is if the holes, initially paired on a plaquette in
isolated ladders, would tend to break the pairs and move
independently on the frustrated chains as the ILC are increased.
There is yet another interesting possibility that is that the
holes form pairs on the frustrated chains. These pairs would have
more kinetic energy than the ones formed on ladders and this
change of pairing would imply a change from a magnetic binding
on ladders to a ``kinetic binding" on chains. In any case, taking
into account the results from the three-chain cluster, we predict
a loosening of the pairing on ladders. It is difficult to answer
these questions by calculations on finite clusters. Exact
diagonalization results for two holes on the $4\times4$ cluster give
support to the above mentioned possibilities. In Fig.~\ref{hh4xn}
the hole-hole correlation functions for $t^\prime=1.0$ at several
distances as a function of $t_{inter}$ are shown. At small
$t_{inter}$, the largest correlation corresponds to a pair of
holes along the diagonal of a plaquette, which is typical of
isolated ladders. Around $t_{inter}=0.7$ there is an abrupt
change to a situation in which the largest correlations correspond
to holes belonging to the same frustrated chain. The second hole
is slightly more likely to be in the other chain of the same
interladder zig-zag chain (site `4' of Fig.~\ref{hh4xn}).
Somewhat smaller is the correlation on the same chain but at the
largest distance on this cluster (site `3').
This behavior is radically different to ladders AF coupled in a
square lattice without frustration. In this case, the d-wave
pair typical of a ladder evolves smoothly to the rotationally
invariant d-wave pair of the square lattice.

\begin{figure}
\begin{center}
\epsfig{file=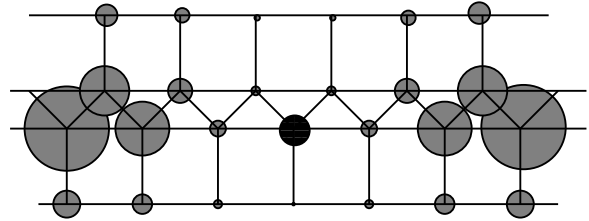,width=8cm,angle=0}
\end{center}
\caption{Most likely hole distribution if a hole is located in the
position indicated by a solid circle. Results obtained on the
$4 \times 6$ cluster for $t^\prime=1.0$ and $t_{inter}=1.0$.
}
\label{hh6x4}
\end{figure}

Similar results are obtained for the $6\times4$ cluster with two
holes. In Fig.\ref{hh6x4}, we show pictorially the most likely
hole probability for $t^{\prime}=t_{inter}=1$. The area of the
circles is proportional to the  probability of finding a hole
if there is a hole in a given site. In this case the largest
probability corresponds to holes located at the maximum distance
along the same chain. The next probability in decreasing order
also corresponds to a hole in the frustrated zig-zag chain. On
this cluster we found that as $t_{inter}$ is increased from zero
the holes initially at a distance $\sqrt{2}$ starts to move away
on the same ladder and finally they move to the same ILC chain.

\begin{figure}
\begin{center}
\epsfig{file=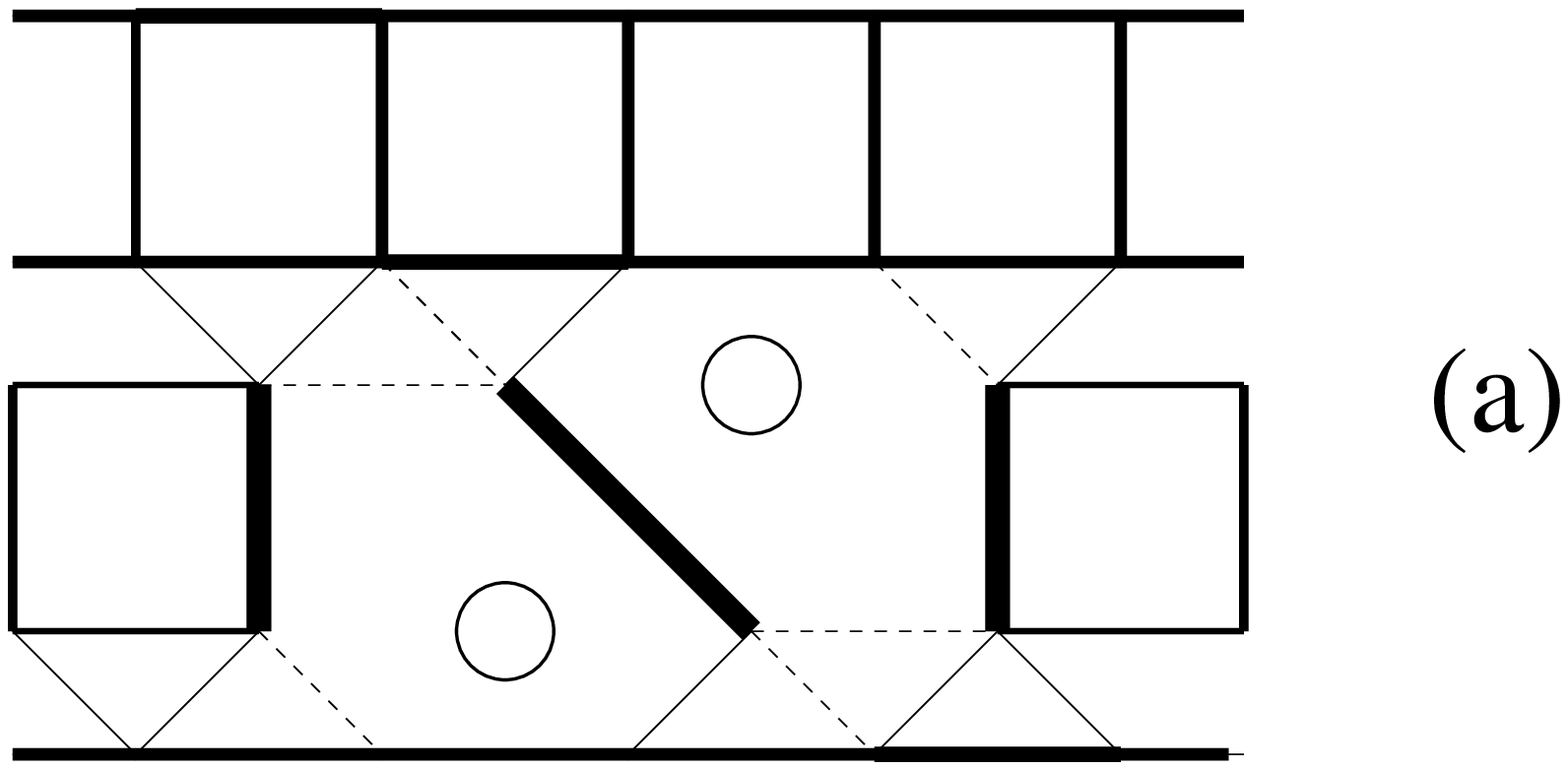,width=6.5cm,angle=0}

\vspace{0.4cm}
\epsfig{file=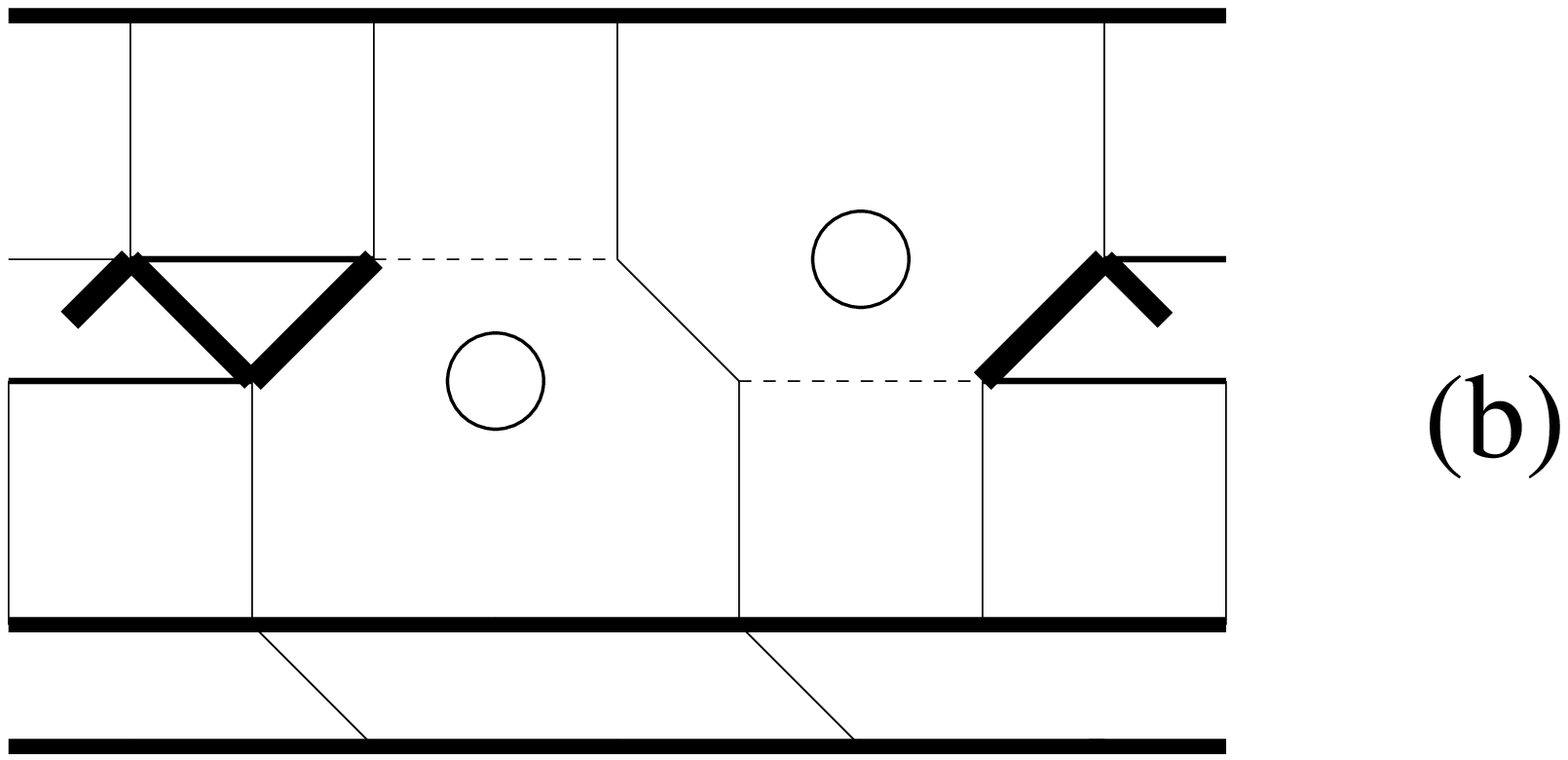,width=6.5cm,angle=0}
\end{center}
\caption{Largest magnetic correlations for the most likely 
hole distribution. The line thickness of each segment is proportional
to the AF correlation of the connected spins. FM correlations are 
indicated with dashed lines. Results obtained on
the $4 \times 4$ cluster for $t^\prime=1.0$, (a) $t_{inter}=0.4$
and (b) $t_{inter}=1.0$.
}
\label{ss4xn}
\end{figure}

Finally, for the sake of completeness, we show in Fig.~\ref{ss4xn}
the largest nearest and next nearest neighbor spin-spin correlations
for small and large values of $t_{inter}$ for the most likely
position of the holes in each case (see Fig.~\ref{hh4xn}). The
structure of these correlations globally agrees with the
schematic picture of Fig.~\ref{esquema}.

\section{Conclusions}

In summary, we have performed numerical studies on strongly
correlated electron systems, as described by the $t-J$ model, on
frustrated coupled ladders, in particular on the trellis lattice.
In the undoped case, QMC simulations, although hampered by the minus
sign problem, allowed us to reach large enough clusters to detect
meaningful changes in the magnetic properties of this system.
In this case, for ladders coupled with frustrating interactions,
we have shown that the peak of the magnetic structure factor
shifts from $(\pi,\pi)$ to $(\pi,\pi/2)$ for low enough temperatures
as the ILC is increased. Moreover, the peak at $(\pi,\pi/2)$ becomes
also the lowest energy excitation. This behavior is very similar
to the one previously found for FM coupled
ladders\cite{dalosto}. We have shown also that this behavior
appears due to the onset of frustration and hence we expect that
it will appear in the trellis lattice as well. This behavior could
be detected experimentally on a ladder compound like
SrCu$_2$O$_3$\cite{azuma} or Sr$_{14}$Cu$_{24}$O$_{41}$ upon a
suitable application of pressure.
In fact, after the submission of this manuscript we
became aware of an experimental study\cite{kiryukhin} on
the similar ladder compound CaCu$_2$O$_3$. The neutron diffraction
results reported in this
manuscript indicate that the magnetic structure is incommensurate
in the direction of the frustrated interladder interaction in the
plane of the ladders, as suggested by the present study (see also
Ref.~\onlinecite{dalosto}).
The validity of the simpler
FM coupled ladders model also explains the vanishing of the spin
gap\cite{mayaffre} as due to increasing interladder couplings.

Next, we have analyzed the evolution of pairing when the ILC are
increased. In this case the physics is governed by short range
effects and so we studied small clusters with exact diagonalization.
Our main result is that ILC suppresses pairing in ladders. Results
on $4\times4$ and $6\times4$ clusters indicate that holes move to the
chains with first and second neighbor interactions formed by the
legs of neighboring ladders and the zig-zag interactions between
them. We have identified the mechanism of this suppression of
pairing as a gain of kinetic energy of the holes by moving on the
frustrated chains. Even for these clusters we have noticed
important size effects which unable us to determine if it appears
an alternative pairing of holes on the frustrated chains or rather
the pairing is completely lost when ILC are large enough and
holes begin to move independently from each other. Finally,
we would like to stress the radically different behavior found
in this case with respect to that found
in ladders coupled in a square lattice, where the d$_{x^2-y^2}$
pairing is preserved.

\acknowledgements

The authors acknowledge many interesting discussions with A.
Dobry, C. Gazza, A. Greco, D. Poilblanc and A. Trumper.
Part of the computations were performed at the Supercomputer
Computations Research Institute (SCRI) and at the Academic
Computing and Network Services at Tallahassee(Florida).


\begin{references}

\bibitem{drs} E.~Dagotto, J. Riera, and D. Scalapino, Phys. 
       Rev. B {\bf 45}, 5744 (1992); T. Barnes, E.~Dagotto, J.
       Riera, and E. Swanson, Phys. Rev. B {\bf 47}, 3196 (1993).

\bibitem{rice93} T. M. Rice, S. Gopalan, and M. Sigrist,
      Europhys. Lett. {\bf 23}, 445 (1993); H. Tsunetsugu, M. Troyer,
       and T. M. Rice, Phys. Rev. B {\bf 49}, 16078 (1994).


\bibitem{noack} R. M. Noack, S. R. White, and D. J. Scalapino, Phys.
     Rev. Lett. {\bf 73}, 882 (1994); R. M. Noack, N. Bulut, D. J.
     Scalapino, and M. G. Zacher, Phys. Rev. B {\bf 56}, 7162 (1997).


\bibitem{uehara} M. Uehara, T. Nagata, J. Akimitsu, N. Mori, and
      K.Kinoshita, J. Phys. Soc. Jpn. {\bf 65}, 2764 (1996).

\bibitem{nagata} T. Nagata, M. Uehara, J. Goto, J. Akimitsu, N. 
     Motoyama, H. Eisaki, S. Uchida, H. Takahashi, T. Nakanishi,
     and N. Mori, Phys. Rev. Lett. {\bf 81}, 1090 (1998).

\bibitem{pachot} S. Pachot, P. Bordet, R. J. Cava, C. Chaillout,
     C. Darie, M. Hanfland, M. Marezio, and H. Takagi,
     Phys. Rev. B {\bf 59}, 12048 (1999).

\bibitem{Note1} It has been noticed that to fit experimental data
     some other couplings should be included in the model, such as an
     interaction along the diagonal of the ladder plaquettes and
     a four-spin interaction on these plaquettes. Y. Mizuno, T.
     Tohyama, and S. Maekawa, Phys. Rev. B {\bf 58}, 14713 (1998);
     S. Brehmer, H.-J. Mikeska, M. M\"uller, N. Nagaosa, and S.
     Uchida, Phys. Rev. B {\bf 60}, 329 (1999).

\bibitem{rpd} J. Riera, D. Poilblanc, and E. Dagotto, Eur. Phys. J. B
      {\bf 7}, 53 (1999).

\bibitem{normandmila} B.~Normand, K. Penc, M. Albrecht, and 
     F.~Mila, Phys. Rev. B {\bf 56}, 5736 (1997).
 
\bibitem{magishi} K. Magishi, S. Matsumoto, Y. Kitaoka, K. Ishida,
    K. Asayama, M. Uehara, T. Nagata, and J. Akimitsu, Phys. Rev. B 
    {\bf 57}, 11533 (1998).

\bibitem{tranquada} J. M. Tranquada, B. J. Sternlieb, J. D. Axe,
     Y. Nakamura, and S. Uchida, Nature {\bf 375}, 561 (1995);
     K. Yamada, C. H. Lee, K. Kurahashi, J. Wada, S. Wakimoto, S. 
     Ueki, H. Kimura, Y. Endoh, S. Hosoya, G. Shirane, R. J. 
     Birgeneau, M. Greven, M. A. Kastner, and Y. J. Kim, Phys.~Rev.
     ~B {\bf 57}, 6165 (1998); N. Ichikawa, S. Uchida, J. M. 
     Tranquada, T. Niemoeller, P. M. Gehring, S.-H. Lee, J. R. 
     Schneider, cond-mat/9910037, and references therein.

\bibitem{white} S. R. White and D. J. Scalapino, Phys. Rev. Lett. 
     {\bf 80}, 1272 (1998); Phys. Rev. B {\bf 61}, 6320 (2000).

\bibitem{miyahara} S. Miyahara, M. Troyer, D. C. Johnston,
      and K. Ueda, J. Phys. Soc. Jpn. {\bf 67}, 3918 (1998).

\bibitem{reger} J. Reger and A. P. Young, Phys. Rev. B {\bf 37},
      5978 (1988), and references therein.

\bibitem{wiese} See e.g., S. Chandrasekharan and U.-J. Wiese,
     Phys. Rev. Lett. {\bf 83}, 3116 (1999), and references
     therein.

\bibitem{haas} See e.g., S.~ Haas, J.~Riera, and E.~Dagotto,
         Phys.~Rev.~B {\bf 48}, 3281 (1993).

\bibitem{dalosto} S. Dalosto and J. Riera, Phys. Rev. B {\bf 62},
       1 July (2000).


\bibitem{eccleston} R. S. Eccleston, M. Uehara, J. Akimitsu, H. Eisaki,
       N. Motoyama, and S. Uchida, Phys.~Rev.~Lett. {\bf 81}, 1702
       (1998).  

\bibitem{whiteaffleck} S. R. White and I. Affleck, Phys.~Rev.~B
       {\bf 54}, 9862 (1996).

\bibitem{Note3} At the classical level, the equation for the pitch
       angle\cite{whiteaffleck} is not modified by the coupling
       between $J_1-J_2$chains.

\bibitem{rieradag} J. Riera and E. Dagotto, Phys. Rev. B {\bf 57},
            8609 (1998).


\bibitem{azuma} M. Azuma, Z. Hiroi, M. Takano, K. Ishida, and Y.
      Kitaoka, Phys. Rev. Lett. {\bf 73}, 3463 (1994).

\bibitem{ogata} Enhancement of pairing was found in the
     $t_1-J_1-J_2$ model at $J_2/J_1=0.5$ by M. Ogata, M. Luchini and T.
     M. Rice, Phys. Rev. B {\bf 44}, 12083 (1991).

\bibitem{mayaffre} H. Mayaffre, P. Auban-Senzier, D. J\'erome, D.
      Poilblanc, C. Bourbonnais, U. Ammerahl, G. Dhalenne, A.
      Revcolevschi, Science {\bf 279}, 345 (1998).

\bibitem{kiryukhin}V. Kiryukhin {\it et al.}, preprint cond-mat/0009158.

\end{references}
\end{document}